\documentclass[aps,preprint]{revtex4}%
\usepackage{amsfonts}
\usepackage{amsmath}
\usepackage{amssymb}
\usepackage{graphicx}%
\begin{document}
\preprint{ }
\title[Limits on gauges]{Strong limitations on allowable gauge transformations in electrodynamics}
\author{H. R. Reiss}
\affiliation{Max Born Institute, 12489 Berlin, Germany}
\affiliation{American University, Washington, DC 20016-8058, USA}
\email{reiss@american.edu}


\begin{abstract}
Conservation principles establish the primacy of potentials over fields in
electrodynamics, both classical and quantum. The contrary conclusion that
fields are primary is based on the Newtonian concept that forces completely
determine dynamics, and electromagnetic forces depend directly on fields.
However, physical conservation principles come from symmetries such as those
following from Noether's theorem, and these require potentials for their
statement. Examples are given of potentials that describe fields correctly but
that violate conservation principles, demonstrating that the correct statement
of potentials is necessary. An important consequence is that gauge
transformations are severely limited when conservation conditions must be
satisfied. When transverse and longitudinal fields are present concurrently,
the only practical gauge is the radiation gauge.

\end{abstract}
\date{27 October 2015}
\maketitle

It is widely accepted that the Aharonov-Bohm effect
\cite{ehrenbergsiday,aharonovbohm} establishes the need for electromagnetic
potentials for the proper description of electromagnetic phenomena, but this
is a unique effect that is specifically quantum-mechanical. It is demonstrated
here that physical conservation principles that exist in both classical and
quantum mechanics are sufficient to establish the primacy of potentials over
fields for the description of electromagnetic phenomena. Basic examples are
shown in which one set of potentials satisfies conservation principles and
another does not, even though both sets of potentials predict exactly the same
fields. A direct consequence is that physically allowable gauge
transformations impose severe limitations on physically valid gauge
transformations. An example from strong-field physics shows that problems
containing both longitudinal and transverse fields can be described only
within the radiation gauge.

The first example is well-known, since it pertains to the description of a
charged particle in a static electric field. This is an energy-conserving
system, consistent with Noether's theorem that the Lagrangian should be
independent of time. However, time-independence is true for only one set of
potentials and not a second one that nevertheless correctly describes the
fields. The second example to be presented is less well-known. It violates the
conservation of the ponderomotive potential energy of a charged particle in a
plane-wave field, and it is only recently that this has been demonstrated
\cite{hr14} to be a conserved quantity. Furthermore, potentials that violate
the conservation rule about the ponderomotive potential are not widely known,
although they have been in the literature \cite{hr79} for many years.

Despite the familiarity of the example of the static electric field, its
implications have not been fully appreciated. The static electric field
$\mathbf{E}_{0}$ can be described by the scalar and vector potentials%
\begin{equation}
\phi=-\mathbf{r\cdot E}_{0},\;\mathbf{A}=0, \label{a}%
\end{equation}
or by the potentials%
\begin{equation}
\widetilde{\phi}=0,\;\widetilde{\mathbf{A}}=-ct\mathbf{E}_{0}. \label{b}%
\end{equation}
(Gaussian units are used throughout.) A charged particle in a constant
electric field is a system that is conservative, which is consistent with the
time-independent potentials (\ref{a}), but the time-dependent potentials
(\ref{b}) are unphysical because Noether's Theorem show that they violate the
energy conservation principle. Starting from the potentials (\ref{a}), it has
been shown \cite{hrjmo} that there is no possible gauge transformation that
can preserve energy conservation. That is, no gauge freedom at all exists for
this case.

The uniqueness of the potentials (\ref{a}) is easily demonstrated. The general
gauge transformation is given by%
\begin{equation}
\widetilde{A}^{\mu}=A^{\mu}+\partial^{\mu}\Lambda,\label{c}%
\end{equation}
where $\Lambda$ is a scalar function that satisfies the homogeneous wave
equation%
\begin{equation}
\partial^{\mu}\partial_{\mu}\Lambda=0.\label{d}%
\end{equation}
The covariant expression (\ref{c}) can be be devolved into scalar and vector
potentials as%
\begin{align}
\widetilde{\phi} &  =\phi+\frac{1}{c}\partial_{t}\Lambda,\label{e}\\
\widetilde{\mathbf{A}} &  =\mathbf{A}-\mathbf{\nabla}\Lambda.\label{f}%
\end{align}
Starting with the potentials (\ref{a}), Eq. (\ref{e}) shows that the necessity
to keep potentials independent of time means that $\Lambda$ can, at most, have
a dependence that is linear in $t$, leading to the form $\Lambda=tf\left(
\mathbf{r}\right)  $. However, this would generate time dependence in
$\widetilde{\mathbf{A}}$. Therefore, dependence of $\Lambda$ is limited to the
form $\Lambda\left(  \mathbf{r}\right)  $. In that case, Eq. (\ref{a})
provides a complete description of the field, meaning that $\Lambda=$
\textit{constant }is the only possibility. This is a trivial result,with the
significance that the potentials (\ref{a}) are the unique potentials for the
description of a constant electric field.

The gauge transformation connecting Eqs. (\ref{a}) and (\ref{b}) is just the
G\"{o}ppert-Mayer gauge transformation \cite{gm} applied in the zero frequency
limit. The unphysical nature of the potentials (\ref{b}) has consequences for
strong-field physics that are explained in Ref. \cite{arxiv}.

The second example relates to the very important case of propagating fields
that may, equivalently, be referred to as transverse fields or plane-wave (PW)
fields. Their profound importance stems from the widespread use of lasers in
laboratory experiments, and lasers produce PW fields. The unique ability of PW
fields to propagate without sources means that the fields that impinge on a
target in a laser experiment can only be a superposition of PW fields. Any
\textquotedblleft contamination\textquotedblright\ by fields other than pure
PW fields damps out over short distances, of the order of wavelengths of the
laser field.

A basic limitation on possible gauge transformations that preserve the
identity of PW fields follows immediately from the premise of special
relativity that all inertial frames of reference are equivalent. This is
expressible by the constraint that the spacetime 4-vector $x^{\mu}$ can occur
only in the Lorentz-scalar combination \cite{schwinger,ss}%
\begin{equation}
\varphi\equiv k^{\mu}x_{\mu}=\omega t-\mathbf{k\cdot r}, \label{g}%
\end{equation}
where $k^{\mu}$ is the propagation 4-vector of the PW field, and $\mathbf{k}$
is the propagation 3-vector. The quantity $\varphi$ is just the phase of a
propagating field. The propagation 4-vector lies on the light cone, meaning
that
\begin{equation}
k^{\mu}k_{\mu}=0. \label{h}%
\end{equation}
A gauge transformation that preserves the basic condition for a PW field is
thus constrained by the limitation that the generator of the gauge
transformation can depend on $x^{\mu}$ only in the form of $\varphi$
\cite{hrjmo,hr14}, meaning that
\begin{equation}
\widetilde{A}^{\mu}=A^{\mu}+\partial^{\mu}\Lambda=A^{\mu}+(\partial^{\mu
}\varphi)\frac{d}{d\varphi}\Lambda=A^{\mu}+k^{\mu}\Lambda^{\prime}, \label{i}%
\end{equation}
where $\Lambda^{\prime}$ is the total derivative of $\Lambda$ with respect to
$\varphi$. A direct consequence of Eq. (\ref{i}) is that%
\begin{equation}
\widetilde{A}^{\mu}\widetilde{A}_{\mu}=A^{\mu}A_{\mu}, \label{j}%
\end{equation}
following from Eq. (\ref{h}) and from the transversality property%
\begin{equation}
k^{\mu}A_{\mu}=0. \label{k}%
\end{equation}
This has the basic physical significance that the ponderomotive potential
$U_{p}$ of a particle of charge $q$ and mass $m$ immersed in a PW field,
defined as%
\begin{equation}
U_{p}\equiv\frac{q^{2}}{2mc^{2}}\left(  -A^{\mu}A_{\mu}\right)  , \label{l}%
\end{equation}
is gauge-invariant under any gauge transformation that preserves the PW
property of the field. It is also, from its definition, Lorentz invariant. The
minus sign in Eq. (\ref{l}) is included to make $U_{p}$ a positive quantity,
since the 4-vector $A^{\mu}$ is a spacelike 4-vector, whose square is negative
in terms of the time-favoring real metric employed here. Two important facts
are that the period average generally employed in the definition of $U_{p}$ is
needlessly limiting, and is not necessary; and that $U_{p}$ defined for a PW
field is a true potential energy \cite{hr14} and not the kinetic
\textquotedblleft quiver energy\textquotedblright\ associated with oscillatory
electric fields.

One of the important properties, just demonstrated in Eq. (\ref{i}), is that
any gauge transformation applied to a PW field is limited to an alteration of
the 4-potential by an additive function lying on the light cone. This
limitation is strong in the sense that adding a light-cone contribution does
not confer any simplifications in a practical problem. If the vector potential
$A^{\mu}$ for the PW field is expressed in the radiation gauge, where
$A^{0}=0$, there is no practical advantage to departing from this simple
choice. The radiation gauge (often called the Coulomb gauge) thus has the
convenient property that transverse fields are represented by 3-vector
potentials, and longitudinal fields are represented by scalar potentials.

There is another reason to limit the possible gauges for the expression of a
PW field to the unique choice of the radiation gauge when a PW field is
applied in the concurrent presence of a longitudinal field, such as a Coulomb
field. When the PW field is very strong, its fundamentally relativistic nature
can be viewed as requiring the entire problem of an atom or molecule in the
strong PW field to be basically relativistic. If the atomic \textquotedblleft
electron\textquotedblright\ in such a problem is viewed as a spinless
particle, as is usually done in AMO physics, then the Klein-Gordon equation is
the appropriate quantum dynamical equation. The Klein-Gordon equation for an
\textquotedblleft electron\textquotedblright\ in a strong laser field was
examined in Ref. \cite{hr90} with the result that a cross-coupling term
between longitudinal and transverse fields was found of the form $2VeA^{0}$,
where $V$ is a Coulomb potential and $\mathbf{A}^{0}$ is the time part of the
4-vector PW potential. This cross coupling would survive into the
nonrelativistic limit, which is the Schr\"{o}dinger equation. The
Schr\"{o}dinger equation does not have such a term, which constitutes a
requirement that the radiation gauge must be selected for the PW field. Stated
briefly: the concurrent existence of a longitudinal field (like the Coulomb
potential) and a transverse field (a laser field) limits gauge selection to
the sole physical possibility of the radiation gauge.

Now the second example will be examined of a gauge transformation that results
in an unphysical set of potentials despite these potentials predicting the
fields correctly and satisfying all of the normal requirements for carrying
out a gauge transformation. To examine this problem, we temporarily set aside
the requirement that $x^{\mu}$ can occur only in the combination $\varphi$ of
Eq. (\ref{g}). The example concerns the fundamentally important propagating
field (or transverse field or PW field). The generating function
\begin{equation}
\Lambda_{K}=-x^{\nu}A_{v}\left(  \varphi\right)  \label{m}%
\end{equation}
is introduced. The subscript \textquotedblleft K\textquotedblright\ is
intended to stand for \textquotedblleft Keldysh\textquotedblright, since the
original introduction of the this transformation \cite{hr79} was in an
(unsuccessful) attempt to place the well-known Keldysh approximation
\cite{keldysh} on a foundation that referred to a laser field rather than to
the oscillatory electric field treated by Keldysh.\ The 4-vector potential so
obtained from the gauge transformation (\ref{c}) produces the result%
\begin{equation}
\widetilde{A}_{K}^{\mu}=-k^{\mu}\left[  x^{\nu}A_{\nu}^{\prime}\left(
\varphi\right)  \right]  ,\label{n}%
\end{equation}
where $A_{\nu}^{\prime}$ refers to the total derivative of $A_{\nu}$ with
respect to $\varphi$. The significance of this result becomes more clear if
the original 4-potential $A^{\mu}$ is in the radiation gauge, in which case
Eq.\ (\ref{n}) can be written as \cite{hr79}%
\begin{equation}
\widetilde{A}_{K}^{\mu}=-\frac{k^{\mu}}{\omega/c}\left[  \mathbf{r\cdot
E}\left(  \varphi\right)  \right]  ,\label{n1}%
\end{equation}
which has the time component%
\begin{equation}
\widetilde{A}_{K}^{0}\equiv\phi_{K}=-\mathbf{r\cdot E}\left(  \varphi\right)
.\label{n2}%
\end{equation}
The covariant expression for the 4-potential (\ref{n}) then appears to be a
relativistic generalization of the ordinary length-gauge scalar potential.
That this is not the case will shortly become clear.

It can be shown readily \cite{hr79} that $\Lambda_{K}$ satisfies the
homogeneous wave equation as specified in Eq. (\ref{d}), \ the 4-vector
potential obtained from the transformation (\ref{m}) satisfies the
transversality condition%
\begin{equation}
k_{\mu}\widetilde{A}_{K}^{\mu}=0, \label{o}%
\end{equation}
and it is also a \textit{Lorentz gauge}, meaning that
\begin{equation}
\partial_{\mu}\widetilde{A}_{K}^{\mu}=0. \label{p}%
\end{equation}
In other words, all of the requirements attached to a gauge transformation are
satisfied. Furthermore, the potentials \ (\ref{n}) produce exactly the correct
PW electric and magnetic fields; perpendicular to each other and perpendicular
to the propagation direction. That is, the gauge transformation is legitimate
in that it satisfies all the usual conditions imposed on gauge transformations
and it reproduces the desired electromagnetic fields exactly.

However, the gauge-transformed 4-vector potential of Eq. (\ref{n}) is unphysical.

One defect is immediately clear: $\widetilde{A}_{K}^{\mu}$ is lightlike rather
than spacelike. In consequence, the importantly gauge-invariant ponderomotive
potential is predicted to be exactly zero in \ the gauge of Eq. (\ref{n}). The
cause of these problems is directly traceable to the fact that the generating
function (\ref{m}) introduces a dependence on the spacetime 4-vector $x^{\mu}$
that is not in the form of $k^{\mu}x_{\mu}$, thus violating a basic premise of
special relativity for the propagation of a light wave.

A dependence on physical reasoning based entirely on the fields and not on
potentials is consequential. An important example comes from Atomic,
Molecular, and Optical (AMO) physics, where it is customary in the description
of laser-produced phenomena to introduce what is generally called the
\textquotedblleft dipole approximation\textquotedblright, viewed in terms of
fields as the statement that the electric field is a function of time only,
with no spatial dependence, and where the magnetic field is neglected
altogether:%
\begin{equation}
\mathbf{E}=\mathbf{E}\left(  t\right)  ,\quad\mathbf{B}=0.\label{q}%
\end{equation}
This is seemingly plausible as long as the wavelength of the field is much
longer than atomic dimensions. However, this criterion can be seriously
misleading when very strong laser fields are present. For example, the fields
(\ref{q}) do not obey the same Maxwell equations as those for propagating
fields \cite{hrmax}, which means that there is no possible gauge connection to
propagating fields (despite the widespread assumption that there is such a
connection). The dipole fields (\ref{q}) have led to the practice of
subjecting both analytical and numerical calculations to the test that
behavior should replicate that of a constant electric field in the
zero-frequency limit. This is a basic premise, for example, in a textbook on
strong-field physics \cite{joachain}, which reflects the
frequently-encountered use of this criterion in the AMO community. However,
from the point of view of potentials, it is clear that the fields of (\ref{q})
are incompatible with the requirements for a propagating field. An electric
field of zero frequency certainly does not have the propagation property, and
the zero-frequency limit of a propagating field is simply an extremely
long-wavelength radio field. Thus, the requirement that there should be a
zero-frequency limit for laser effects that is a constant electric field is
fundamentally unphysical.

Another widespread concept is that laser-caused atomic ionization is a
tunneling process wherein the attractive Coulomb potential of an electron
bound in an atom can be wholly or partially compromised by the addition of a
scalar potential representing an oscillatory electric field. This is a process
that is completely dependent on the superposition of two longitudinal fields
and cannot be representative of the interaction of a transverse field with a
Coulomb potential. The tunneling model has some success, but it is plainly
confined to a limited domain of frequencies and intensities of the laser
field. In particular, the \textquotedblleft tunneling limit\textquotedblright%
\ where the Keldysh parameter%
\begin{equation}
\gamma_{K}=\sqrt{E_{B}/2U_{p}} \label{r}%
\end{equation}
( where $E_{B}$ is the field-free binding energy of the electron) becomes
extremely small ($\gamma_{K}\ll1$), is actually a domain where the magnetic
field becomes important, the fields of Eq. (\ref{q}) do not properly represent
the laser environment, and conclusions based on the tunneling concept are
inappropriate \cite{hr101,hrtun}.

The principal conclusions reached above can be summarized:

\# Electromagnetic fields are more fundamental than the fields that can be
derived from them in both classical and quantum electrodynamics.

\# There exist potentials that predict fields correctly but violate physical
conservation laws.

\# Gauge transformations that connect physically possible gauges are severely limited.

\# For each of the important special cases of constant electric fields and of
laser fields (i.e., propagating fields) there exists only one physical gauge.

\# Methods based on the electric dipole approximation of AMO physics can
support unphysical conclusions about strong laser fields.

\end{document}